\documentclass[a4]{epl2}
\usepackage{graphicx}
\usepackage{epsfig}
\usepackage{color}
\title{Consequences of complex Lorentz force and  violation of Lorenz gauge condition}
\shorttitle{Consequences of complex Lorentz force and  violation of Lorenz gauge condition}

\author{A. I. Arbab\inst{}\footnote{Present Address. aiarbab@uofk.edu}}
\institute{\inst{a}Institute for Condensed Matter Theory
Department of Physics
University of Illinois at Urbana-Champaign,
1110 West Green Street, IL 61801, USA, and\\
  \inst{b} Department of Physics,
Faculty of Science, University of Khartoum, P.O. Box 321, Khartoum
11115, Sudan
}

\abstract{The complex Lorentz force is introduced and extended to include magnetic scalar. This scalar is found to be associated with a prevailing magnetic field permeating the whole space. It also introduce an extra force in Lorentz complex force. The magnetic scalar is  associated with the vacuum energy. The Proca-Maxwell's massive electrodynamics
is derived from the extend current-density transformations.  Proca-Maxwell's theory is found to be invariant under the extended gauge transformations (current-charge density).  The Lorenz gauge condition is shown to  express the photon charge conservation.  Any violation of Lorenz gauge (photon charge) or electronic charge conservation would lead to spin zero scalar particles. This is manifested in superconductivity. The total charge comprising the electron and photon is always conserved. Owing to superconductivity, the photon charge is related to electron charge by $e_p=\sqrt{\frac{m_p}{m_e}}\,\,e$. Photons inside superconductors are shown to  be massive. It is shown that Maxwell's equations expressed in complex form are more convenient to study duality transformations. }

\begin{document}
\baselineskip=20pt
\voffset=-1.5cm
%\large
\maketitle
\section{Introduction}
Maxwell's equations are the basic equations for studying all electromagnetic interactions \cite{1}. They are invariant under  Lorentz and gauge transformations. They entail that the photon is massless. In de Broglie hypothesis any micro-particle is accompanied with wave nature. The photon on the other hand exhibits particle nature. Hence, duality is the fundamental nature existing for all micro-particles. Accordingly, electrons and photons have dual nature.

Proca extended Maxwell's theory to include massive photons \cite{proca}. The addition of the mass term in the  Maxwell's Lagrangian would break the gauge invariance. The loss of this invariance would make the theory less predictive. However, Proca theory is Lorentz invariant but not gauge invariant. Gauge invariance is applied only to the electric and magnetic fields but not to the current and charge densities that are main ingredient of Maxwell theory. Maxwell's theory guarantees the conservation of charge. This is expressed by the continuity equation relating the spread of charge and current densities. Maxwell's equations are sometimes expressed in terms of vector and scalar potentials instead of electric and magnetic fields. If we redefine these potentials, Maxwell's equations are still unchanged. We say that Maxwell's equations are gauge invariant. The question that will arise whether these potentials are physical observables or mere mathematical constructs. It is demonstrated experimentally by Bohm and Aharonov that these potentials are physical objects \cite{bohm}. To solve Maxwell's equations in terms of potentials the Lorenz gauge is generally adopted, though other gauges are allowed. Hence, Lorenz gauge must have some physical aspect. The unification of quantum mechanics and Maxwell's equations resulted in quantum electrodynamics. Here the electron and photon must interact with each other. The gauge particle, i.e., the photon, is responsible to mediate the interaction between electrons (or charged particles). It is called gauge boson. It has spin equals to 1. While the electron is charged, the photon is left uncharged (neutral). Thus, the photon is chargeless and massless in the standard formulation of quantum electrodynamics. In Proca-Maxwell's theory the photon is massive but no charge is associated to it. However, in Bardeen -Cooper- Schrieffer theory, the superconductivity is described as a microscopic effect caused by a condensation of Cooper pairs into a boson-like state \cite{super}. In this theory two electrons bind together to form Cooper pairs with spin equals to zero. London related the supercurrent to photon field (the vector potential) \cite{lond}. In this sense, the gauge transformations would alter this supercurrent. We should then think to restore the original picture. To this end we should assume that gauge transformations include current density as well as vector potential. Thus, the charge-current densities and scalar-vector potentials should be equally treated.

In this paper we study the consequences of employing the extended complex Maxwell equations.
We introduced in section 2 the generalized Maxwell's equations. We proposed a magnetic scalar field that can give rise to longitudinal waves besides the ordinary electromagnetic transverse waves. This scalar is found to be connected with the charge conservation.

In section 3 we introduced an extended Lorentz force employing quaternions. In this section we obtained an extended Lorentz force embodying the new magnetic scalar. With this representation we obtain the complex Lorentz force relating the force on magnetic and electric charges placed in an electromagnetic field. Besides, we obtained the power associated with these charges. We further show that the extended Lorentz force is duality invariant. In section 4 we showed that when no electromagnetic field is present the magnetic scalar induces a force on the charges present. We associate the energy with this state to that of the vacuum. This energy is reflected in the amount of vacuum energy found today as prescribed by the cosmological constant riddle.

In section 5 we study a complex magnetic scalar and its effect on duality transformation of Maxwell's equation and Lorentz force. In section 6 we relate the Lorenz gauge condition to conservation of photon charge. We showed here that the violation of Lorenz gauge or charge conservation leads immediately to massive photons. We compare the formulation of Proca with that of the  the extended complex Maxwell's formulation. We employ here the London's ansatz for the superconductivity and relate the photon mass with photon charge. This relation agrees with the recent proposition by Chu that gravity (mass) resulted from the non-neutrality of photons \cite{chu}. The charge and mass of the photon  are exceedingly small, but non-zero.

\section{The generalized Maxwell's equations}
We have shown recently that one can generalize Maxwell's equations to include scalar (longitudinal) wave \cite{cpb}. This is achieved by expressing Maxwell's equations in quaternionic form employing an electromagnetic  vector
$\vec{F}=\frac{\vec{E}}{c}+i\vec{B}$ \cite{cpb}. This yields
\begin{equation}\label{1}
\widetilde{\nabla}^*\widetilde{F}=\mu_0\widetilde{J}\,,
\end{equation}
where
\begin{equation}
\widetilde{F}=(\Lambda\,, i\vec{F})\,,\qquad\widetilde{J}=(i\rho\,c\,, \vec{J})\,,\qquad
\widetilde{\nabla}^*=\left(\frac{i}{c}\frac{\partial}{\partial t}\,,-\vec{\nabla}\right)\,,
\end{equation}
and the scalar $\Lambda$ defines some scalar 'magnetic' function representing the fourth component of the electromagnetic 4-vector. Using the quaternionic multiplication rule for two quaternions, $\tilde{A}=(a_0\,,\vec{a})$ and $\tilde{B}=(b_0\,,\vec{b})$, one finds, $\tilde{A}\tilde{B}=\left(a_0b_0-\vec{a}\cdot\vec{b}\,,\,\,a_0\vec{b}+\vec{a}\,b_0+\vec{a}\times\vec{b}\right)$.
Hence, the expansion of eq.(1), using eq.(2), yields
\begin{equation}\label{1}
\frac{1}{c}\frac{\partial \Lambda}{\partial t}+\vec{\nabla}\cdot\vec{F}=\mu_0\,c\rho\,,
\end{equation}
and
\begin{equation}\label{1}
-\vec{\nabla}\Lambda-\frac{1}{c}\frac{\partial \vec{F}}{\partial t}-i\vec{\nabla}\times\vec{F}=\mu_0\vec{J}\,,
\end{equation}
where  $\mu_0$ and $\varepsilon_0$  are  the permeability and permitivity of the free space, respectively, and $c^{-2}=\varepsilon_0\mu_0$.
Equations (3) and (4) generalize Maxwell's equations to incorporate scalar wave, $\Lambda$.
Taking the divergence of eq.(4),  differentiating eq.(3) partially with respect to time, and adding the two resulting equations, one finds
\begin{equation}
\frac{1}{c^2}\frac{\partial^2\Lambda}{\partial t^2}-\nabla^2\Lambda=\mu_0\left(\vec{\nabla}\cdot\vec{J}+\frac{\partial\rho}{\partial t}\right)\,.
\end{equation}
It is interesting to see that $\Lambda$ doesn't influence the electromagnetic wave nature.
\section{The extended Lorentz force}
The extended Lorentz  complex force can be written as
\begin{equation}\label{1}
\tilde{f}=q\tilde{V}\tilde{F}^*\,,\qquad \tilde{V}=\left(ic\,, \vec{v}\right)\,.
\end{equation}
Using the quaternion multiplication rule, eq.(2) and eq.(6) yield
\begin{equation}\label{1}
\tilde{f}=\left(iq(c\Lambda+\vec{v}\cdot\vec{F})\,,\,\, q\Lambda\vec{v}+qc\vec{F}-iq\vec{v}\times\vec{F}\right)\,.
\end{equation}
The vector part of the quaternion in eq.(7) is the extended Lorentz  complex force. Hence,
\begin{equation}\label{1}
\vec{f}_\Lambda= q\Lambda\vec{v}+qc\vec{F}-iq\vec{v}\times\vec{F}\,.
\end{equation}
This force reduces to the ordinary complex Lorentz  force for magnetic and electric charges
\begin{equation}\label{1}
\vec{f}_L= qc\vec{F}-iq\vec{v}\times\vec{F}\,.
\end{equation}
It is important to point that we have derived the complex Lorentz force in eq.(9) from Maxwell's equations \cite{cpb}.
Now, eq.(8) generalizes eq.(9) to include  magnetic scalar, $\Lambda$. This scalar affects both the field and force equations. It is interesting to see whether eq.(8) can be obtained from eqs.(3) and (4) following the same steps that led to eq.(9) \cite{cpb}.
The generalized Lorentz force is associated with the force on electric and magnetic charges.  It is  generally associated with the symmetrised Maxwell's equations. But in the present formulation we didn't presume a priori the presence of magnetic charges.  It seems that the magnetic charge is associated naturally with the electric charge. This is evident if we express the total charge of a particle as, $q=q_e-i\frac{q_m}{\mu_0c}$. Equation (8) is thus the extended complex Lorentz force that is associated with the complex Maxwell's equations above.

The scalar part of eq.(7) represents the power dissipated by  moving charges in the electric and magnetic fields, \emph{i.e.},
\begin{equation}\label{1}
P=iqc^2\Lambda+iqc\vec{v}\cdot\vec{F}=-qc\vec{v}\cdot\vec{B}+iq(c^2\Lambda+\vec{v}\cdot\vec{E})\,.
\end{equation}
Equations (8) and (10) can be written as
\begin{equation}\label{1}
\vec{f}_\Lambda= q(\Lambda\vec{v}+\vec{E}+\vec{v}\times\vec{B})+iqc(\vec{B}-\frac{\vec{v}}{c^2}\times\vec{E})\,.
\end{equation}
and
\begin{equation}\label{1}
P_\Lambda=-qc\vec{v}\cdot\vec{B}+i(qc^2\Lambda+q\vec{v}\cdot\vec{E})\,.
\end{equation}
Equation (9) can be written as
\begin{equation}\label{1}
\vec{f}_L= q(\vec{E}+\vec{v}\times\vec{B})+iqc(\vec{B}-\frac{\vec{v}}{c^2}\times\vec{E})\,.
\end{equation}
We have  recently derived from Maxwell's equations the equation \cite{cpb}
\begin{equation}
q\frac{d\vec{F}}{dt}=\vec{\nabla}\times\vec{f}_D\,,
\end{equation}
where
\begin{equation}
\vec{f}_D=-q\vec{v}\times\vec{F}-iqc\vec{F}
\end{equation}
Under duality transformations, $\vec{F}\rightarrow -i\vec{F}$, so that eq.(15) transforms to eq.(9). Thus, $\vec{f}_D$ is the dual of $\vec{f}_L$. Furthermore, the invariance of eq.(8) under  the duality transformations dictates that a real  $\Lambda$ must vanish. Now the invariance of the complex Lorentz force ($\vec{f}_L=m\vec{a}$), eq.(9), under the duality transformation ($\vec{F}\rightarrow-i\vec{F}$)  requires that either\\
case (a):$$m\rightarrow -i\, m\,\qquad {\rm and}\qquad q\rightarrow q\,,$$
or\\
case (b):
$$q\rightarrow i\, q\,\qquad {\rm and}\qquad m\rightarrow m\,.$$
 Case (a) suggests that one can write the mass as a complex quantity, consisting of the mass of an electric charge ($m_e$) and the mass of a magnetic charge ($m_m)$. That means,
$$m=m_e+i\,m_m\,,$$
so that under duality transformation, $m_e\rightarrow m_m$ and  $m_m\rightarrow -m_e$. Case (b) suggests also, $q=q_e-i\frac{q_m}{\mu_0c}$\,, so that under duality transformation, $\mu_0cq_e\rightarrow -q_m$ and  $q_m\rightarrow \mu_0cq_e$. If we consider case (a), then Lorentz force is invariant under duality transformation. Similarly, Lorentz force is invariant under the duality transformation suggested in case (b). Case (a) conforms with Maxwell's field equations where, $\rho\rightarrow -i\rho$,\footnote{$\rho=\rho_e+i\frac{\rho_m}{\mu_0c}$.} \emph{i.e.},  $\mu_0c\rho_e\rightarrow\rho_m$ and $\rho_m\rightarrow-\mu_0c\rho_e$. However, case (b) needs some attention. It will be effective for the quantum electrodynamics theory where the particle and the fields are involved. It is interesting to remark that we have dealt with case (b) where we propose the invariance of a theory under the complex space-time transformation of massive particles \cite{arb-wid}, generalizing t'Hooft-Nobenhuis transformation \cite{tooft}.

Now eqs.(11) and (12) can be expressed as
\begin{equation}\label{1}
\vec{f}_\Lambda= q_e(\Lambda\vec{v}+\vec{E}+\vec{v}\times\vec{B})+iq_m(\vec{H}-\vec{v}\times\vec{D})\,.
\end{equation}
and
\begin{equation}\label{1}
P_\Lambda=-q_m\vec{v}\cdot\vec{H}+iq_e(c^2\Lambda+\vec{v}\cdot\vec{E})\,,
\end{equation}
where $q=q_e$ and $q_m=q_ec\mu_0$ are the electric and magnetic charges, $\vec{B}=\mu_0\vec{H}$, $\vec{D}=\varepsilon_0\vec{E}$.
It is pertinent to mention that the first derivation of the Lorentz force was due to Oliver Heaviside in 1889, but  Lorentz derived it afterwards.
However, the Lorentz force in eq.(13) describes the force not only on electric charge but on magnetic charge too. This is so despite the fact that we didn't employ the symmetrised Maxwell equation in our exposition. The electric charge gains energy  from the electric field, \emph{i.e.}, $q_e\vec{v}\cdot\vec{E}$, abd the magnetic scalar present in space.  Moreover, we observe that $\Lambda$ influences the force on the particle and the power associated with it.
\section{The vacuum contribution}
Note that in the absence of electric and magnetic fields and their sources, there exists a non-zero force acting on the charge, \emph{viz.}, $\vec{f}_0=q_e\Lambda\vec{v}$, where $\Lambda$ is now constant, and the  power delivered  is, $P_0=q_ec^2\Lambda$. Thus, the total force acting on a neutral system is zero. Note that when $\Lambda$ is constant, the extended Maxwell's equations reduce to the ordinary Maxwell's equations. The Lorentz force in this case looks like a frictional (viscous drag) force that a particle  will experience when moves in space.  However, no force can act on the magnetic charge when no fields are present. This force can be attributed to the vacuum (ether) that acts like a fluid (magnetic) in which the charge moves. Therefore, the speed of  a negative charged particle (an electron),  decreases exponentially with time as $v\propto e^{\frac{q_e\Lambda}{m}t}$, and the distance traveled is $x=\frac{m}{q_e\Lambda}\, v+x_0$, where $x_0$ is some constant. Since this force is proportional to $\Lambda $, which has a dimension of magnetic field, we can ascribe this force to a relic (background) magnetic field prevailing the whole space that is unidirectional. A moving charged particle will be separated away from its antiparticle by the ether. This may help account for the non-annihilation of all matter in during the early cosmic evolution.  To have an idea of this magnetic field one can define a characteristic time, $\tau_\Lambda=\frac{m}{q_e\Lambda}$, that reflects the properties of this field. If we equate for instance, $\tau_\Lambda$ and $m$ to Planck's time ($10^{-43}\rm \,s$) and mass ($10^{-8}\,\rm kg$), respectively, then $\Lambda_P$ amounts to, $\Lambda_P\sim 10^{53}T$. Similarly, for the present epoch ($10^{18}\,\rm s$), one has $\Lambda_0\sim 10^{-8}T$. These two values are obtained considering that the universe is a quantum system \cite{quantum1, quantum2}. This prevailing magnetic field may arise from the charge the photon carries. Hence,  the magnetic scalar $\Lambda$ reflects the level to which the electric charge is violated. This scalar field may also be associated with the magnetic field carried by the photons in the cosmic background radiation. This is possible if we  associate  a charge with the photon. The smallness of $\Lambda$ today signifies that even if the photon does have a charge, its value is exceedingly small.  At any rate, we should expect that $f_0$ to be very small. When the universe was created $f_P\sim 10^{42}N$. This coincides with the maximal  force existing in nature, i.e., $\frac{c^4}{4G}$, where $G$ is the gravitational constant \cite{quantum1, quantum2}. The maximum present acceleration ($a_0=\frac{q_e\Lambda\,c}{m}$) that can act by the ether on a charged Planck's mass amounts to $10^{-10}\,\rm ms^{-2}$. This coincides with the acceleration exhibited by Pioneer 10/11 acceleration anomaly \cite{ander}. We therefore think of the space as a fluid endowed with intrinsic magnetic properties. The maximum energy embedded in the ether is $E_0=q_e\Lambda\,c\,\Delta\, t$. Thus, the energy absorbed by the ether during the entire cosmic expansion is equal to the one absorbed during Planck's time ($10^9\rm\, J$). The mass creation rate is defined by $\dot m=e\Lambda$. Hence, the mass created since Planck's time is $M=e\Lambda\,\Delta\,t\sim 10^{53}\rm\, kg$, where $\Delta\,t\sim 10^{18}\rm\,sec$, and $\Lambda\sim 10^{53}\rm\,T$. The corresponding power also coincides with the maximal power,  \emph{i.e.}, $\frac{c^5}{G}$ \cite{quantum2}. In this case one can write $\Lambda= \frac{c^3}{q_eG}$. This gives the magnetic field produced by a charged particle moving in gravitational field. Thus, a magnetic field can originate from gravity. The energy associated with $\Lambda$ will act as a vacuum energy in the universe. It seems that the magnetic field induced at the time of the universe creation (big bang) is still not yet fully extracted from the space. The prevalence of magnetic scalar in our universe mimics the phenomenon of hysteresis in magnetic material. This ushers that space has magnetic nature retaining the time with it.

Let us now calculate the magnetic flux produced in a system of area $A$. If this flux is quantized, one can write, $A=\frac{\tau_\Lambda\hbar}{m}$. This shows that the area containing the magnetic flux is quantized. For instance for the whole universe today, $A_0=10^{52}\rm m^2$, where $\hbar\sim 10^{87}\rm J\,s$, $m\sim 10^{53}\rm kg$, and at Planck's time, $A_P=10^{-70}\rm m^2$ \cite{quantum1, cosmic}. These are of the same order of magnitudes for these values. Furthermore, one finds the rate of decay of the magnetic flux is constant. This  means the induced electromotive force during the time it is developed is constant throughout cosmic expansion. The vacuum energy density associated with this magnetic scalar is $\rho_\Lambda=\frac{\Lambda^2}{2\mu_0}$. This yields a value of $10^{-10}\,Jm^{-3}$ for the present epoch. Hence, one can associate the cosmological constant ($\Lambda_v$) with this magnetic scalar \cite{wein}. This can be written as $\Lambda_v=\frac{G}{c^2k}\,\Lambda^2$, where $k$ is the Coulomb's constant. It is interesting to see that the ratio, $\frac{\Lambda_v}{\Lambda^2}=\frac{G}{c^2k}$ remains constant. Thus, the equality between the two contributions, one due to $\Lambda_v$ and the other due to $\Lambda$, which is negative, can lead to a zero total contribution solving the cosmological constant dilemma. Furthermore, the above ratio gives the induced magnetic capacitance per mass. If we consider our universe to be spherical of radius $R$, then this capacitance, $C=4\pi\varepsilon_0R$. Hence, $R=\frac{Gm}{c^2}$. This is of the same order as the Schwarzschild radius of a gravitating object.
\\
Because today monopoles (magnetic charges) are not abundantly observed, duality symmetry is a broken symmetry. Particle physicists can associate an energy scale at which duality symmetry is broken. Notice that  a real magnetic scalar breaks the duality invariance, but a complex one preserves the symmetry. Thus, the occurrence of $\Lambda$ implies a duality breaking.

 The effective temperature associated with a uniformly accelerating observer ($a$) in  vacuum  is given by \cite{unruh}
 $$a=\frac{2\pi c k_BT}{\hbar},$$
 where $k_B$ is the Boltzman constant.  If we assume the origin of this acceleration to the one mentioned above, then
 $$\Lambda=\frac{2\pi mk_BT}{e\hbar}\,,$$
 where we set $q_e=e$, representing the remnant (vacuum) magnetic field. The corresponding energy density associated with this magnetic field is thus
 $$u_\Lambda=\left(\frac{2\pi^2k_B^2m^2}{\mu_0e^2\hbar^2}\right)\, T^2\,.$$
 This may provide the vacuum energy density associated with cosmological scale. Note that the energy density of an ordinary boson gas (photon) is related to the temperature by $u\propto T^4$. It is interesting to note that Gasperini related the cosmological constant to $T^2$, thus the cosmological constant can be interpreted
as a parameter measuring the intrinsic temperature of the empty space \cite{gasper}. Thus, because of $\Lambda$ space and time are tightened together.

 The magnetic scalar can be written as
  $$\Lambda=\frac{4\pi^2}{\lambda^2} \left(\frac{\hbar}{e}\right)\,,\qquad \lambda=\frac{h}{\sqrt{2\pi mk_BT}}\,.$$
 where $\lambda$ is the wavelength (thermal) of the system at which quantum effects are effective, and\, $\hbar/e$ \, is the  quantum flux. Hence, $\Lambda$ can be related to the magnetic field of the emitted black body radiation. The mass $m$ can be related to the mass of the photon.

 The force associated with this vacuum is thus
 $$F=\frac{2\pi mc k_BT}{\hbar}\,.$$
 If we equate this force to the maximal force in nature ($c^4/4G$), we obtain the maximal temperature
  $$T=\frac{c^3\hbar}{8\pi Gm k_B}{}\,.$$
This is exactly the Hawking temperature of a black hole radiating like  a black body radiation \cite{hawking}. Note that the  vacuum  in  the  uniformaly accelerated  frame  is  supposed to be in  thermal equilibrium  with  its  own radiation. Therefore, the above Hawking relation is a manifestation of quantum gravity and thermodynamics. It is not necessarily connected with black holes. The maximal force occurs when the gravitational energy is converted into relativistic mass energy. In this case pair-production is created. While the mass $m$ in our above relation is the mass of the newly created particles from the conversion of gravitational energy into mass energy, the mass in Hawking is the mass of the evaporated black hole. Thus, if the Planck mass, $10^{-8}\rm kg$, was converted into radiation, its temperature would be $10^{31}\rm K$. And if the whole present universe mass, $10^{53}\rm kg$, is converted to radiation, its temperature will be $10^{-29}\rm K$. Therefore, the former temperature represents the temperature of the vacuum filling the primeval universe (big bang), and the latter one is the temperature of the present vacuum.
 It is shown that in a strong magnetic field the vacuum behaves as a superconductor, and similarly at very low magnetic field \cite{super}. The magnetic field was so enormous at the time of big bang, and exceedingly small at the present time.

\section{The complex magnetic scalar}
If we now express $\Lambda$ as a complex function, \emph{viz.}, $\Lambda=\Lambda_m+\frac{i}{c}\Lambda_e$,\,\footnote{This because $\Lambda$ has a dimension of magnetic field.} \cite{cpb}, then eqs.(11) and (12) read
\begin{equation}\label{1}
\vec{f}_\Lambda= q_e\left(\Lambda_m\vec{v}+\vec{E}+\vec{v}\times\vec{B}\right)+iq_m\left(\varepsilon_0\Lambda_e\vec{v}+\vec{H}-\vec{v}\times\vec{D}\right)\,.
\end{equation}
and
\begin{equation}\label{1}
P_\Lambda=-q_m\left(\frac{\Lambda_e}{\mu_0}+\vec{v}\cdot\vec{H}\right)+iq_e\left(c^2\Lambda_m+\vec{v}\cdot\vec{E}\right)\,.
\end{equation}
Hence, a complex $\Lambda$ scalar affects the electric and magnetic charges. It is worth to mention that the magnetic charge is analogous to electric charge, but is accelerated by a magnetic field, and its path is bent by an electric field. With such behavior, magnetic charges can be detected.

Let us now consider the duality transformations, \emph{i.e.},
\begin{equation}\label{1}
\vec{E}(\vec{cB})\rightarrow c\vec{B}(-\vec{E})\,,\,\,\, c\Lambda_m(\Lambda_e)\rightarrow \Lambda_e(-c\Lambda_m)\,,\,\,\, c\mu_0q_e(q_m)\rightarrow q_m(-c\mu_0q_e)\,,
\end{equation}
so that $\vec{f}_\Lambda\rightarrow\vec{f}_\Lambda$, but $P_\Lambda\rightarrow -P_\Lambda$. This resembles the transformation of force and power under time reversal. Does this mean the duality transformation is equivalent to time reversal elsewhere?

Now for Maxwell's equations, eq.(5), to be invariant under duality, $\vec{J}(\rho)\rightarrow -i\vec{J}(-i\rho)$, and  $\Lambda\rightarrow -i\Lambda$ . This implicitly implies that $\vec{J}$ and $\rho$ are complex quantities too. This is indeed the case if we define $\vec{J}=\vec{J}_e+i\frac{\vec{J}_m}{c\mu_0}$ and $\rho=\rho_e+i\frac{\rho_m}{c\mu_0}$. In this case, the ordinary, eqs.(5), and extended, eqs.(3) \& (4), Maxwell's equations are invariant under the duality transformations.
Therefore, it is more convenient to study duality transformations involving electric and magnetic charges and their belongings, if we express Maxwell's equation in complex form rather than the ordinary real form.
The energy conservation of the electromagnetic field can be written as
\begin{equation}
\frac{\partial}{\partial t} \frac{1}{2\mu_0} (\vec{F}\cdot\vec{F}^*)+\vec{\nabla}\cdot\frac{ic}{2\mu_0}(\vec{F}\times\vec{F}^*)=-\vec{J}\cdot\vec{F}^*\,.
\end{equation}
This equation is invariant under the duality transformation.
\section{The charge-current densities transformations}
Charge conservation dictates that the right hand-side of eq.(5) to vanish. Hence, the scalar function $\Lambda$
satisfies the wave equation. Thus, besides the electromagnetic field which  has a transverse nature, a scalar
field with a longitudinal character is predicated from the generalized  Maxwell's equations, \emph{viz.},
eqs.(3) and (4). These equations can be obtained from the ordinary Maxwell's equations if relax the Lorenz
gauge condition to read \cite{analog},
\begin{equation}
\vec{\nabla}\cdot\vec{A}+\frac{1}{c^2}\frac{\partial\varphi}{\partial t}=-\Lambda\,.
\end{equation}
Equations (3) and (4) can be expanded to yield
\begin{equation}\label{1}
\vec{\nabla}\cdot\vec{E}=\frac{\rho}{\varepsilon}_0-\frac{\partial \Lambda}{\partial t}\,,\qquad
\vec{\nabla}\cdot\vec{B}=0\,,
\end{equation}
and
\begin{equation}\label{1}
\vec{\nabla}\times\vec{E}=-\frac{\partial \vec{B} }{\partial t}\,,\qquad
\vec{\nabla}\times\vec{B}=\mu_0\vec{J}+\frac{1}{c^2}\frac{\partial \vec{E}}{\partial t}+\vec{\nabla}\Lambda\,.
\end{equation}
Notice that $\Lambda$ is a real (physical) scalar wave since it has energy and momentum. Moreover, note that
$\Lambda$ has a dimension of magnetic field.

Equations (23) and (24) mimic the vacuum solution if we let
\begin{equation}\label{1}
\frac{\rho}{\varepsilon}_0=\frac{\partial \Lambda}{\partial t}\,,\qquad\mu_0\vec{J}=-\vec{\nabla}\Lambda\
\end{equation}
These two equations state that
\begin{equation}\label{1}
\frac{\partial \rho}{\partial t}+\vec{\nabla}\cdot\vec{J}=\frac{1}{c^2}\frac{\partial^2 \Lambda}{\partial t^2}-\nabla^2\Lambda=0\,,
\end{equation}
and
\begin{equation}\label{1}
\vec{\nabla}\rho+\frac{1}{c^2}\frac{\partial \vec{J}}{\partial t}=0\,,\qquad\vec{\nabla}\times\vec{J}=0\,.
\end{equation}
Hence, charge is conserved as long as $\Lambda$ satisfies the wave equation. It is interesting that this case corresponds to our generalized continuity equations \cite{wida}.
Equations (26) and (27) are invariant under the following charge-current gauge transformations
\begin{equation}\label{1}
\vec{J}\,'=\vec{J}+\vec{\nabla}\beta\,,\qquad \rho\,'=\rho-\frac{1}{c^2}\,\frac{\partial \beta}{\partial t}\,,
\end{equation}
where $\beta$ is some scalar function satisfying the wave equation.
Equations (23) and (24) can be obtained by applying the transformations in eq.(28) in Maxwell's equations, where $\Lambda=\mu_0\beta$.
If we now allow charge non-conservation, in particular we consider
\begin{equation}
\vec{\nabla}\cdot\vec{J}+\frac{\partial\rho}{\partial t}=-\frac{\mu^2}{\mu_0}\,\Lambda\,,
\end{equation}
where $\mu=\frac{mc}{\hbar}$ and $m$ is a mass scale, then eq.(26) yields
\begin{equation}\label{1}
\frac{1}{c}\frac{\partial^2 \Lambda}{\partial t^2}-\nabla^2\Lambda+\mu^2\Lambda=0\,.
\end{equation}
This is the Klein-Gordon equation of spin zero particles. Thus, the violation of charge conservation or Lorenz gauge condition lead to an
emergence of massive scalar particle. If this scalar is associated with the photon, then the photon would have
three polarization sates instead of two. But if it is linked to a different particle then one can associate this
to Higgs boson. Thus, photons inside superconductors can develop a non-zero  mass. Equations (22) and (29) suggest the following relations
\begin{equation}
\vec{J}=-\alpha\,\vec{A}\,,\qquad \rho=-\frac{\alpha}{c^2}\,\varphi \,,\qquad \alpha=\frac{\mu^2}{\mu_0}\,.
\end{equation}
Under gauge transformations (scalar-vector potentials) one has
\begin{equation}\label{1}
\vec{A}\,'=\vec{A}+\vec{\nabla}\lambda\,,\qquad \varphi\,'=\varphi-\frac{\partial \lambda}{\partial t}\,.
\end{equation}
The  definition in eq.(31) is the one considered by London in his theory of superconductivity, with $\alpha=ne^2/m$
\cite{lond}.
Equation (31) suggests also the following transformations in Maxwell's theory
\begin{equation}
\vec{J}\rightarrow\vec{J}-\alpha\,\vec{A}\,,\qquad \rho\rightarrow\rho-\frac{\alpha}{c^2}\,\varphi \,.
\end{equation}
that can lead to massive photon. This is indeed the case, since the application of eq.(33) in Maxwell's equations
yields the Proca-Maxwell's electrodynamics for massive photon \cite{proca}. These are
\begin{equation}\label{1}
\vec{\nabla}\cdot\vec{E}=\frac{\rho}{\varepsilon_0}-\mu^2\varphi\,,\qquad \vec{\nabla}\cdot\vec{B}=0\,.
\end{equation}
and
\begin{equation}
\vec{\nabla}\times\vec{E}=-\frac{\partial \vec{B}}{\partial
t}\,,\,\,\,\,\vec{\nabla}\times\vec{B}=\mu_0\vec{J}+\frac{1}{c^2}\frac{\partial\vec{E}}{\partial t}-\mu^2\vec{A}
\,.
\end{equation}
Equations (34) \& (35) are equivalent to eqs.(23) \& (24) if we let
\begin{equation}\label{1}
\mu^2\varphi=\frac{\partial\Lambda}{\partial t}\,,\qquad -\mu^2\vec{A}=\vec{\nabla}\Lambda\,,
\end{equation}
which satisfy the Lorenz gauge condition
\begin{equation}\label{1}
\vec{\nabla}\cdot\vec{A}+\frac{1}{c^2}\frac{\partial\varphi}{\partial t}=0\,.
\end{equation}
It is worth to mention that the Proca-Maxwell's theory, that is not invariant
under the normal gauge transformations, is now invariant under the matter-field  transformations, eq.(33),
including both $\vec{J}\,(\vec{A})$ as well as  $\rho\,(\varphi)$  as dictated by eqs.(28) \& (32), where $\beta=\alpha\,\lambda$ (or $\Lambda=\mu^2\lambda$).
The mystery of the scalar $\Lambda$ is still worth further consideration. At any rate one can associate $\Lambda$ with the amount by which the charge is violated or to the mass of the photon. It is also related to the gauge scalar, $\lambda$.

\section{Electron-photon interaction}
In standard quantum  electrodynamics the photon is considered to be chargeless and massless. In this section we would like to explore the consequences that the photon has both mass and charge.  In this sense the photon carries energy and charge.
Adding eqs.(22) and (29) using eq.(33), one finds
\begin{equation}
\vec{\nabla}\cdot\vec{J}_T+\frac{\partial\rho_T}{\partial t}=0\,,
\end{equation}
where
\begin{equation}
\vec{J}_T=\vec{J}+\vec{J}_p\,,\qquad \rho_T=\rho+\rho_p\,,
\end{equation}
where
$\vec{J}_p=-\alpha\vec{A}$ and $\rho_p=-\alpha\varphi/c^2$, are the current and charge densities of photons. This urges us to interpret the Lorenz gauge condition as a manifestation of the conservation of photonic charge, or possibly magnetic charge. If the photon had mass it would be possible to have charge.
Dirac however, associated a magnetic charge $(q_m$) with the electric charge ($q_e$) quantization \cite{dirac},
\begin{equation}
q_m\,q_e=n\hbar/2\,,
\end{equation}
where  $n$ is an integer.
This relation can be seen as expressing the amount of charge violating. This is associated with the Heisneberg's uncertainty in determining the electric and magnetic charges simultaneously, \emph{viz.}
\begin{equation}
\Delta q_m\,\Delta q_e=\hbar/2\,,
\end{equation}
It is interesting to see that under duality invariance, the quantization condition in eq.(40) requires that $n\rightarrow -n$.
Equations (38) and (39) reveal that the total charge of the system is conserved.
We further assume that the constant $\alpha$ is the same for photon and electron, and that the photon number density ($n_p$) is equal to the electron number density ($n_e$). This implies that the photon charge is
\begin{equation}
e_p=\sqrt{\frac{m_p}{m_e}}\,\, e\,.
\end{equation}
This equation agrees with the relation suggested recently by Chu  assuming that gravity has a purely electromagnetic origin embedded in the non-neutrality of photons. He proposed that $m=Cq^2$, for a particle with mass $m$, charge $q$ and the constant $C$ is related to the gravitational constant and is equal to $C=3.55\times 10^7\,\rm kgC^{-2}$~\cite{chu}.  He obtained a value of\,\, $e_p=7\times 10^{-20}e$. Equation (42) relates the photon charge to its mass.
Using the black-body radiation distribution, we have recently deduced that the photon mass today is $\sim 10^{-68}\rm kg$~ \cite{quantum1, quantum2}. Hence, eq.(42) states that the photon charge would be
\begin{equation}
e_p\sim 10^{-19}\, e\,.
\end{equation}
This is in a remarkable agreement with Chu proposition. If we now
connect the baryon-to-photon ratio ($10^{-10}$) to equality of
electronic charge to photonic charge, \emph{i.e.},
$N_p\,e_p=N_e\,e$, assuming charge neutrality, thus $N_e/N_p=e_p/e$.
Then,  $e_p/e=10^{-10}$ so that eq.(42) yields $m_p\sim 10^{-50}\rm
kg$. The photon charge-mass relationship depends on what photon we
are studying. Thus, this ratio may differ from one case to another.
An upper limit for the photon charge is found recently by
Semertzidis \emph{et al.} from Laser light deflection in magnetic
field to be \, $ 8.5\times10^{-17}\,e$ ~\cite{laser}.

If we now equate the ratio of the electric force  to the gravitational force between two photons to the ratio between the vacuum energy at Planck's time to the one at the present time (i.e., $\sim 10^{120}$), then one finds $e_p/m_p\sim 10^{50}$~\cite{quantum1, quantum2}. This result is compatible with eq.(40).
If we consider now that photons are accumulated inside superconductors  with rest energy density equals to that of a black-body radiation, $\rho_\gamma=\frac{\pi^2}{15}\frac{(k_BT)^4}{(c\,\hbar)^3}=n_p\,m_p\,c^2$, where $k_B$ is the Boltzman constant,  then one finds $m_p\sim 10^{-61}\,\rm kg$ for a temperature of few Kelvins. For  a comprehensive limit on the photon mass we suggest that the reader should refer to reference \cite{mass}.
\section{Concluding remarks}
We have investigated the consequences of of using Complex Lorentz force and abandoning the Lorenz gauge condition in formulating electromagnetic theory.  Several physical effects would arise as result of enlarging the Maxwell formulation.

\end{document}